\documentclass[prb,twocolumn,showpacs,amsmath,amssymb,amstext,amsfont]{revtex4}
\usepackage{amsmath}
\usepackage{graphicx}
\usepackage{bm}
\usepackage[usenames]{color}
\usepackage{verbatim}
\newcommand{\be}{\begin{equation}}
\newcommand{\ee}{\end{equation}}   
\newcommand{\bea}{\begin{eqnarray}}
\newcommand{\eea}{\end{eqnarray}}
\newcommand{\phrl}[1]{Phys.~Rev.~Lett. {\bf #1}}
\newcommand{\phrb}[1]{Phys.~Rev.~B {\bf #1}}

\newcommand{\jpsj}[1]{J.~Phys.~Soc.~Jpn. {\bf #1}}
\newcommand{\RMP}[1]{Rev.~ Mod.~ Phys. {\bf #1}}
\newcommand{\bib}{\bibitem}

\newcommand{\lp}{\left(}
\newcommand{\rp}{\right)}

\renewcommand{\k}{\mathbf{k}}

\begin{document}

\title{Domain walls in superconductors: Andreev bound states and tunneling features}

\author{S. P. Mukherjee and K. V. Samokhin}
\affiliation{Department of Physics, Brock University, St. Catharines, Ontario, Canada L2S 3A1}

\begin{abstract}
Domain walls can be formed in superconductors with a discrete degeneracy of the ground state, for instance, due to the breaking of time reversal symmetry. 
We study all cases where the formation of domain walls is possible in a tetragonal superconductor with the point group $D_{4h}$. 
We discuss both triplet and mixed singlet order parameters. It is found that in all cases the domain walls support subgap Andreev bound states, whose energies 
strongly depend on the direction of semiclassical propagation. The bound state contribution to the density of quasiparticle states exhibits peculiar features, which can be observed in tunneling experiments. 

\end{abstract}


\maketitle

\section{Introduction}
\label{sec:I}

The order parameter inhomogeneities in unconventional superconductors and superfluids, such as the Abrikosov vortices, twin and grain boundaries, interfaces, magnetic impurities, and 
domain walls (DWs) are the breeding ground for gapless localized quasiparticles with distinct topological properties.\cite{Qi} On the other hand, the very presence of DWs is a direct evidence of an unconventional nature of the pairing, 
because the DWs can appear only in a superconductor with two or more distinct degenerate ground states, which transform one into another by some discrete symmetry operations, e.g. by time reversal. 
This is possible if the superconducting order parameter has more than one component, i.e. corresponds to either a multi-dimensional irreducible representation of the crystal point group or to 
a mixture of different, e.g., one-dimensional (1D), representations. 

One of the best-studied examples of superconductors or superfluids which can support DWs is a spin-triplet chiral $p$-wave state, which might be realized in Sr$_2$RuO$_4$ (Ref. \onlinecite{Mackenzie}) or in 
thin films of superfluid ${}^3$He-$A$ (Ref. \onlinecite{Vol92}). The order parameter in the chiral $p$-wave state corresponds to the odd two-dimensional (2D) representation, $E_u$, of the tetragonal point group $D_{4h}$ and 
is proportional to $k_x\pm ik_y$. DWs separate the opposite chirality states, $k_x+ik_y$ and $k_x-ik_y$, which break time reversal symmetry (TRS) and have the same energy in the absence of external magnetic field. 
One possible mechanism of the DW formation is that domains of opposite chirality appear spontaneously in different parts of the system upon cooling across the phase transition, due to the sample inhomogeneity. 
Alternatively, an increase in the gradient energy near the DW might be compensated by the creation of low-energy quasiparticles bound to the DW, which is particularly effective in one-dimensional systems.\cite{KY02} 
Experimentally, there is evidence of DWs in Sr$_2$RuO$_4$, both from the magnetic field modulation of the single-face Josephson critical current\cite{Kidwin06} and from the anomalous hysteresis of 
voltage-current characteristics,\cite{Kambara08} and also in slabs of superfluid ${}^3$He (Ref. \onlinecite{WWG04}). Nonchiral states with the order parameters proportional to $k_x\pm k_y$ or $k_x,k_y$ 
are also phenomenologically possible for $p$-wave pairing.\cite{Mineev} In both cases the ground state is twofold degenerate and, therefore, DWs can exist. 

Other examples of superconducting systems which admit the existence of DWs are $d+is$ and $d+id$ states. In these cases the pairing is singlet and predominantly $d$-wave, with an admixture of a subdominant component of a different symmetry, 
$s$-wave or another $d$-wave, in the bulk.
The possibility of such fully gapped states, which break both time reversal and a discrete lattice symmetry, had been first discussed theoretically for high-$T_c$ cuprates,\cite{KotRokh} but it was later ruled out based on 
phase-sensitive tests of pairing symmetry.\cite{TK00} Recently, there has been a renewal of interest in the $d+is$ states in the context of iron-based superconductors, both theoretically\cite{dpis-Fe-based} and experimentally.\cite{Tafti14}  
Various other TRS-breaking states have also been discussed for URu$_2$Si$_2$ (Ref. \onlinecite{Schemm}), Ba$_{1-x}$K$_x$Fe$_2$As$_2$ (Ref. \onlinecite{spis}), UPt$_3$ (Ref. \onlinecite{UPt3}), 
doped graphene (Ref. \onlinecite{NLC}), undoped bilayer silicene (Ref. \onlinecite{Liu2}), SrPtAs (Ref. \onlinecite{Fischer}), Na$_x$CoO$_2\cdot y$H$_2$O (Ref. \onlinecite{KPHT}), and PrOs$_4$Sb$_{12}$ (Ref. \onlinecite{AC07}).

Motivated by the growing evidence of the existence of multiply degenerate, in particular TRS-breaking, states in unconventional superconductors, in this paper we study how the DWs separating different ground states affect the spectrum
of fermionic quasiparticles. The order parameter variation across a DW creates a quasiparticle bound state, resulting in characteristic features in the density of states (DOS), which can be detected, e.g., in tunneling measurements.
We focus on the quasi-2D tetragonal case, which is applicable to both Sr$_2$RuO$_4$ and the iron-based materials. While most of the previous works considered only the DW bound state spectrum for the chiral $p$-wave state,\cite{p-wave-ABS,Sam12} 
we also study the $d+is$ and $d+id$ states, as well as nonchiral $p$-wave states. While we consider mixed pairing states in the bulk, another possibility is that a subdominant order parameter is stabilized only near a surface, 
in which case the spectrum of the surface bound states was found in Ref. \onlinecite{RBFS98}.  
The paper is organized as follows.
In Sec. II,  we derive a general expression for the DW bound state energy in the semiclassical (Andreev) approximation. In Sec. III, we list all possible cases which admit the DW formation for the point group $D_{4h}$. 
In Secs. IV and V, we discuss the triplet and mixed singlet cases, respectively, and calculate the spectrum of the bound states as well as their contribution to the quasiparticle DOS. Finally, we summarize our results in Sec. VI. 
Throughout the paper we use the units in which $\hbar=e=c=1$.

\section{Andreev bound states}
\label{sec:II}

We consider a quasi-2D superconductor with a cylindrical Fermi surface and assume that the DW is along the $y$-axis, so that the superconducting domains are extended on both sides of the DW to infinity along the $x$-axis. 
The order parameter variation takes place within a region of width $\xi_d$ near the DW.
Throughout our derivation we use the Andreev approximation,\cite{And64,AGSY99} in which the Bogoliubov quasiparticles propagate along semiclassical trajectories characterized by the Fermi-surface
wavevectors $\k_F=(k_{F,x},k_{F,y})=k_{F}(\cos\theta,\sin\theta)$. This approximation is justified since the scales of variation of the superconducting order parameter, including the DW width $\xi_d$, are much greater than
the Fermi wavelength $k_F^{-1}$. 

\begin{figure}
\includegraphics[width=2.0in,height=2.0in, angle=0]{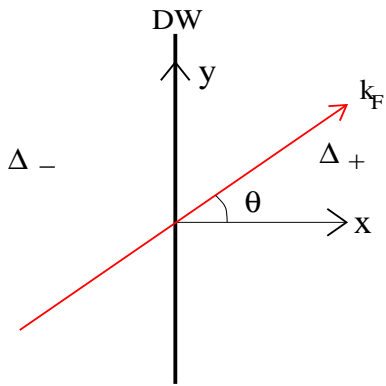} 
\caption{(Color online) Schematic diagram showing the DW, the semiclassical trajectory of a Bogoliubov quasiparticle, and the order parameters $\Delta_{\pm}$ on 
both sides.}
\label{fig:trajectory}
\end{figure}

For a given direction of semiclassical propagation, the gap function takes the form $\Delta_{\k_F}(x)=\Delta(\theta,x)$, with the $\theta$-dependence determined by 
the pairing symmetry, see Sec. \ref{sec:III} below.
The unavailability of an exact analytical expression for the $x$-dependence of the DW order parameter leads to various approximation schemes, see Refs. \onlinecite{Volovik} and \onlinecite{p-wave-DWs} for the chiral $p$-wave case.  
We use a simple, analytically treatable, model in which the DW is assumed to be a sharp boundary of zero width located at $x=0$. Then the gap function has the following form:
\begin{equation}
\label{Delta-pm-definition}
  \left.\begin{array}{ll}
        \Delta_{\k_F}(x)=\Delta_{+}(\theta),& x>0, \medskip \\ 
        \Delta_{\k_F}(x)=\Delta_{-}(\theta),& x<0,
        \end{array}\right.
\end{equation}
i.e. it is characterized by two different complex numbers along each semiclassical trajectory, see Fig.~\ref{fig:trajectory}.

In the Andreev approximation, the quasiparticle wave function is a product of a rapidly oscillating plane wave $e^{i\k_F\bm{r}}$ and a slowly varying envelope function 
$\Psi(x)$, which satisfies the equation
\be
\lp\begin{array}{cc} -iv_{F,x}\nabla_{x} & \Delta_{\k_F}(x)\\ \Delta^*_{\k_F}(x) & iv_{F,x}\nabla_{x} \end{array}\rp \Psi(x) = E\Psi(x),
\label{Andreev}
\ee
where $v_{F,x}=v_F\cos\theta$ and $v_F$ is the Fermi velocity.
Using the sharp DW model, Eq. (\ref{Delta-pm-definition}), we find that the bound-state wave function has the form $\Psi_\pm(x)\sim e^{\mp\kappa_{\pm}x}$ at $x>0$ ($x<0$), where 
$\kappa_{\pm}=\Omega_{\pm}/|v_{F,x}|$, with $\Omega_{\pm}=\sqrt{|\Delta_{\pm}|^2-E^2}\geq 0$. 
In the general pairing case, the envelope wave function has four components, but in all the models we consider below the spin channels decouple and the equations can be reduced to a two-component form. 

Substituting the wave functions into the boundary condition $\Psi(-0)=\Psi(+0)$, we arrive at the following equation for the bound state energy:
\be
\frac{\Delta_{+}}{E - i v_{F,x}\kappa_{+}} =\frac{\Delta_{-}}{E + i v_{F,x} \kappa_{-}},
\ee
which can be represented in the form
\begin{equation}
\label{tilde-E-eq}
\tilde{E} + i \Omega_{-}=\gamma(\tilde {E} - i \Omega_{+}),
\end{equation}
where $\tilde{E}=E\,\mathrm{sgn}(v_{F,x})$ and $\gamma=\Delta_{-}/\Delta_{+}=\gamma_R+i\gamma_I$. Taking the real and imaginary parts of Eq. (\ref{tilde-E-eq}), we obtain:
\begin{equation}
\label{Omega-pm}
  \Omega_{+}=\frac{1-\gamma_R}{\gamma_I}\tilde{E},\qquad \Omega_{-}=\frac{|\gamma|^2-\gamma_R}{\gamma_I}\tilde{E}.
\end{equation}
The first of these equations gives
\be
\label{tilde E}
\tilde {E} = \mathrm{sgn}(\tilde {E}) \frac{1}{\sqrt{1+\beta^2}}|\Delta_{+}|,
\ee
where
\begin{equation}
\label{beta-def}
  \beta=\frac{1-\gamma_R}{\gamma_I}.
\end{equation}
The sign function that appears on the right-hand side of Eq. (\ref{tilde E}) can be assigned using the condition that $\Omega_+\geq 0$, from which we find
$\mathrm{sgn}(\tilde {E})=\mathrm{sgn}(\beta)$. Therefore, we finally obtain for the Andreev bound state (ABS) energy:
\be
  E_b(\theta)=|\Delta_{+}(\theta)|\frac{1}{\sqrt{1+\beta^2(\theta)}}\,\mathrm{sgn}[\beta(\theta)\cos\theta].
\label{E_bound}
\ee
There is an additional constraint coming from the second of the expressions (\ref{Omega-pm}). Since $\Omega_{-}\geq 0$, we have
\be
\mathrm{sgn}(|\gamma|^2-\gamma_R)\,\mathrm{sgn}(1-\gamma_R)=1.
\label{cond}
\ee
Thus, for each direction of semiclassical propagation satisfying the condition (\ref{cond}), there is one ABS with the energy given by Eq. (\ref{E_bound}). For singlet pairing, the gap function is even in momentum and 
$\Delta_\pm(\theta+\pi)=\Delta_\pm(\theta)$, while for triplet pairing, the gap function is odd and $\Delta_\pm(\theta+\pi)=-\Delta_\pm(\theta)$. It follows from Eqs. (\ref{beta-def}) and (\ref{E_bound}) that 
$\beta(\theta+\pi)=\beta(\theta)$ and, therefore, $E_b(\theta+\pi)=-E_b(\theta)$ in all cases. 

The ABS wave function is exponentially localized on both sides of the DW, with the characteristic scales given by $\kappa_{\pm}^{-1}(\theta)=|v_{F,x}|/\sqrt{|\Delta_{\pm}|^2-E_b^2}$. 
The sharp DW approximation is valid for those directions of semiclassical propagation for which the DW width $\xi_d$ is smaller than $\kappa_{\pm}^{-1}$. 
This condition is strongly angle-dependent and, in particular, fails for ``grazing'' trajectories, for which $\theta$ is close to $\pm\pi/2$, therefore, $v_{F,x}\to 0$ and the Andreev approximation itself is inapplicable. 
However, for most directions of $\bm{k}_F$, one can use the following rough estimate: $\kappa_{\pm}^{-1}\gtrsim v_F/\Delta_0\sim\xi$, where $\Delta_0$ is a characteristic value of the gap and $\xi$ is the superconducting correlation length. 
Therefore, the sharp DW model is quantitatively justified if $\xi_d\lesssim\xi\lesssim\kappa_{\pm}^{-1}$.

The ABS's contribute to the quasiparticle DOS, which can be probed in tunneling experiments. The momentum along the $y$-axis is conserved and the bound-state DOS per unit length of the DW
is given by the following expression:
$$
  N_b(E)=\int_{-k_F}^{k_F}\frac{dk_{F,y}}{2\pi}\sum_{k_{F,x}}\delta[E-E_b(\k_F)],
$$
where $k_{F,x}=\pm\sqrt{k_F^2-k_{F,y}^2}$. Making the change of variables $k_{F,x}=k_F\cos\theta$, $k_{F,y}=k_F\sin\theta$, we obtain:
\be
\label{DOS-general}
  N_b(E)=N_Fv_F\int_0^{2\pi}d\theta\,|\cos\theta|\delta[E-E_b(\theta)],
\ee
where $N_F$ is the Fermi-surface DOS in two dimensions.

\section{Nonuniform states in a tetragonal superconductor}
\label{sec:III}

The point group $D_{4h}$ has ten irreducible representations in total (considering even as well as odd representations), of which eight are 1D and two are 2D.
The formation of DWs is possible only for those superconducting classes which are degenerate with respect to some discrete symmetry.\cite{Volovik,Mineev} We consider the cases of singlet and triplet 
pairing separately.

Triplet order parameter is described by a spin vector ${\mathbf d}(\k)$. We assume that the spin-orbit coupling fixes the direction of ${\mathbf d}(\k)$ along 
the $z$ axis. Since the 1D representations cannot support DWs, we consider the odd irreducible representation $E_{u}$, which corresponds to $p$-wave pairing. 
The semiclassical gap function takes the following form:
\be
\label{Delta-pwave-general}
\Delta_{\k_F}(x)= \eta_{1}(x)\hat k_{F,x} + \eta_{2}(x)\hat k_{F,y},
\ee
where $\bm{\eta}=(\eta_1,\eta_2)$ is the two-component order parameter, which transforms like a vector in the $xy$ plane, and the basis functions of $E_u$ are given by $\hat k_{F,x}=\cos\theta$, $\hat k_{F,y}=\sin\theta$. 
In Table \ref{tab:D4h} we show all stable states for which the formation of DWs is possible. 
Note that only the states with $\bm{\eta}\propto(1, \pm i)$ break TRS and do not have gap nodes in the bulk, while the other two states preserve TRS but break some discrete crystal symmetries and have bulk gap nodes.

\begin{table}
\caption{\label{tab:D4h} Superconducting states corresponding to the irreducible representation $E_u$ of the point group $D_{4h}$ with 
strong spin-orbit coupling. The first column lists the possible superconducting classes, where $R$ stands for time reversal (the notations are the same as in Ref. \onlinecite{Volovik}), the 
second and third columns show the number of degenerate ground states and the specific forms of the order parameter in the two domains, 
and the vector ${\mathbf d}(\k)$ is given in the last column.}
\begin{tabular}{|c|c|c|c|}
\hline 
$D_{4h}$&
$\mathrm{degeneracy}$&
$\bm{\eta}$&
$\mathbf{d}(\k)$ \tabularnewline
\hline
\hline 
$D_{4}(E)$&
$2$&
$(1,i)$&
$\hat{z}(k_x + i k_y)$  \tabularnewline
&
&
$(1,-i)$ &
$\hat{z}(k_x - i k_y)$   \tabularnewline
\hline 
$D_{4}(C_{2})\times R$&
$2$&
$(1,0)$&
$\hat{z}k_x$  \tabularnewline
&
&
$(0,1)$ &
$\hat{z}k_y$   \tabularnewline
\hline 
$D_{2}(C_{2}')\times R$&
$2$&
$(1,1)$&
$\hat{z}(k_x + k_y)$  \tabularnewline
&
&
$(1,-1)$ &
$\hat{z}(k_x - k_y)$   \tabularnewline
\hline 
\end{tabular}
\end{table}

In the singlet case and quasi-2D geometry (the latter implies $k_z$-independent basis functions), TRS breaking and DWs are possible only in the states formed by a combination of two pure states belonging to two different 1D 
irreducible representations.\cite{Wenger} The semiclassical gap function can be written as
\begin{equation}
\label{Delta-singlet-gen}
  \Delta_{\k_F}(x)=\psi_1(x)\varphi_1(\hat{\k}_F)+\psi_2(x)\varphi_2(\hat{\k}_F),
\end{equation}
where $\psi_{1,2}$ are the order parameters and $\varphi_{1,2}$ are the basis functions of the 1D representations, satisfying $\varphi_{1,2}(\hat{\k}_F)=\varphi_{1,2}(-\hat{\k}_F)$. 
We consider the following three possibilities: a mixture of $B_{1g}$ and $B_{2g}$ representations, for which
\be
\label{varphi-dd}
  \varphi_1=\hat{k}_{F,x}^2 - \hat{k}_{F,y}^2,\qquad \varphi_2=2\hat{k}_{F,x} \hat{k}_{F,y},
\ee
a mixture of $B_{1g}$ and $A_{1g}$ representations, for which
\be
\label{varphi-d2s}
  \varphi_1=\hat{k}_{F,x}^2 - \hat{k}_{F,y}^2,\qquad \varphi_2=1, 
\ee
and a mixture of $B_{2g}$ and $A_{1g}$ representations, for which
\be
\label{varphi-dxys}
  \varphi_1=2\hat{k}_{F,x} \hat{k}_{F,y},\qquad \varphi_2=1.
\ee
The superconducting state in the bulk is given by $(\psi_1,\psi_2)\propto(\Delta_1,\pm i\Delta_2)$, where $\Delta_1$ and $\Delta_2$ characterize the strengths of the individual pairing components. 
The first of the above states is referred to as a $d_{x^2-y^2}\pm id_{xy}$ state, while the last two are $d_{x^2-y^2}\pm i s$ and $d_{xy}\pm is$ states, respectively. 
All of these states are fully gapped in the bulk.\cite{A-2g} 

\begin{figure}
\includegraphics[width=2.5in,height=4.5in, angle=0]{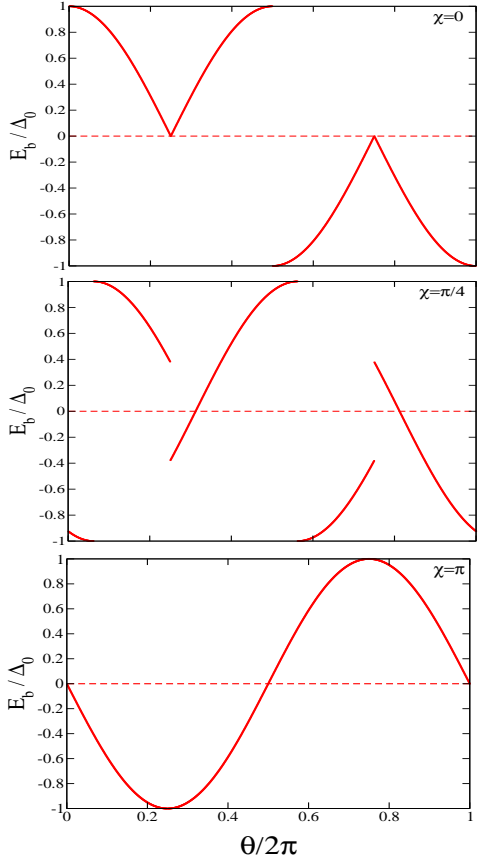}
\caption{(Color online) The variation of the ABS energy with the direction of the quasiparticle propagation for domains with $\bm\eta\propto(1,\pm i)$, for different values of $\chi$.}
\label{fig:EC}
\end{figure}
\vspace*{0.2in}
\begin{figure}
\includegraphics[width=2.5in,height=5in, angle=0]{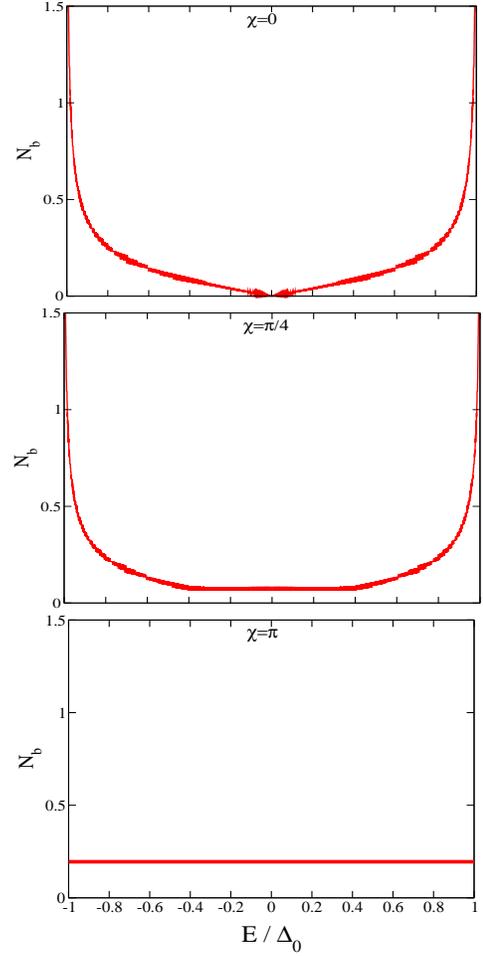}
\caption{(Color online) The energy dependence of the ABS contribution to the DOS (in arbitrary units) for domains with $\bm\eta\propto(1,\pm i)$, for different values of $\chi$.}
\label{fig:DOSC}
\end{figure}

\section{Triplet states}
\label{sec:IV}

In this section we discuss the ABS spectra and the corresponding DOS for the three triplet $p$-wave states listed in Table \ref{tab:D4h}.

\subsection{$\bm{\eta}\propto(1,\pm i)$}

In the chiral $p$-wave state, the order parameters on both sides of the DW have the following form:
\begin{equation}
\label{eta-chiral}
  \left.\begin{array}{ll}
        \bm{\eta}=\Delta_0(1,i),& x<0,\\ 
        \bm{\eta}=\Delta_0(1,-i)e^{i\chi},& x>0,
        \end{array}\right.
\end{equation}
where $\Delta_0$ is the gap magnitude and $\chi$ is the common (Josephson) phase difference between the domains. The latter has to be included in order to satisfy the current conservation across the DW. Its value
depends on the microscopic details of the system, see Appendix A. Without loss of generality we assume $0\leq\chi\leq\pi$.

It follows from Eqs. (\ref{Delta-pm-definition}), (\ref{beta-def}), and (\ref{Delta-pwave-general}) that $\Delta_{-}(\theta)=\Delta_0 e^{i\theta}$, $\Delta_{+}(\theta)= \Delta_0 e^{i \chi} e^{-i\theta}$, 
and
$$
  \beta=\tan\left(\theta-\frac{\chi}{2}\right).
$$
Note that the condition (\ref{cond}) is automatically satisfied, since $|\gamma|^2-\gamma_R=1-\gamma_R$, and
Eq.~(\ref{E_bound}) can be used for all $\theta$ and $\chi$. For the ABS energy we then obtain:
\be
  \frac{E_b(\theta)}{\Delta_0}=\cos\left(\theta-\frac{\chi}{2}\right) \mathrm{sgn}\left[\cos\theta\sin\left(\theta-\frac{\chi}{2}\right)\right],
\ee
see also Ref. \onlinecite{Sam12}. In Fig.~\ref{fig:EC} we plot this last expression for three representative values of the common phase difference $\chi$. The ABS energy has discontinuities at the special directions of semiclassical propagation: 
at $\theta=\pm\pi/2$, i.e. for the quasiparticles moving parallel to the DW, in which case the Andreev approximation is not applicable, and also at 
$\theta=\chi/2$ and $\theta=\chi/2+\pi$, for which $\Delta_+=\Delta_-$ and, therefore, the DW is ``invisible'' to the quasiparticles. 

Next, we calculate numerically the ABS contribution to the DOS per unit length of the DW by using Eq. (\ref{DOS-general}) and plot it in Fig.~\ref{fig:DOSC}. 
One notable feature of them is the presence of van Hove singularities at the gap edge, i.e. at 
$E_b=\pm\Delta_0$ for $\chi\neq\pi$. These singularities are due to the vanishing of the slope of $E_b(\theta)$ at $\theta=\chi/2$ and $\theta=\chi/2+\pi$.
For $\chi=\pi$ the slope vanishes for $\theta=\pi/2,3\pi/2$, where $\cos\theta=0$ and the van Hove singularities are absent. In fact, in the latter case, the DOS is independent of energy.

\begin{figure}
\includegraphics[width=2.5in,height=4.5in, angle=0]{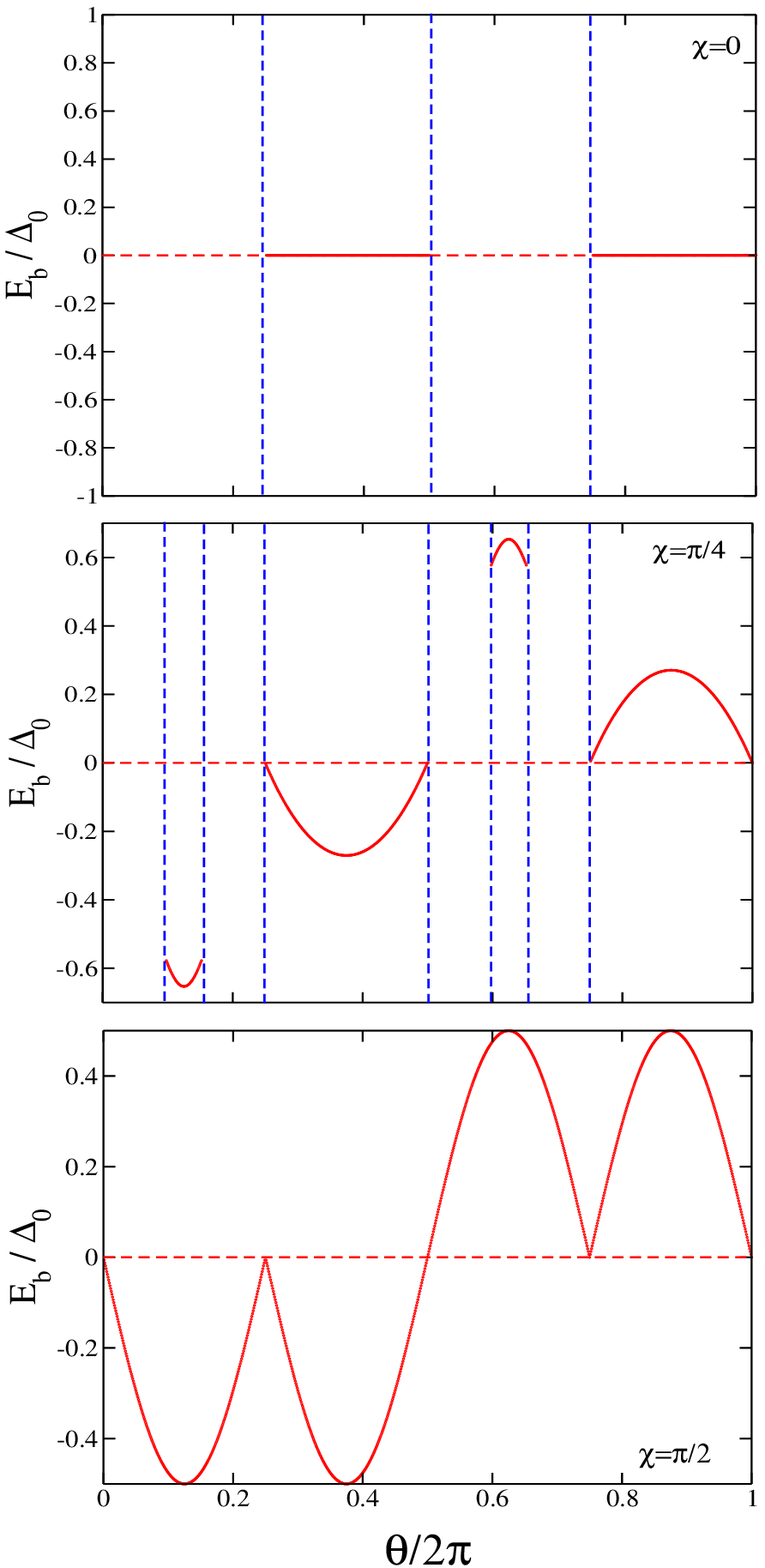}
\caption{(Color online) The variation of the ABS energy with the direction of the quasiparticle propagation for domains with $\bm\eta\propto(1,0)$ and $(0,1)$, for different values of $\chi$. 
Regions bounded by blue vertical dotted lines and not containing any bound states correspond to the directions of propagation for which the condition (\ref{cond}) is not satisfied.}
\label{fig:E1}
\end{figure}
\vspace*{0.2in}
\begin{figure}
\includegraphics[width=2.5in,height=5.0in, angle=0]{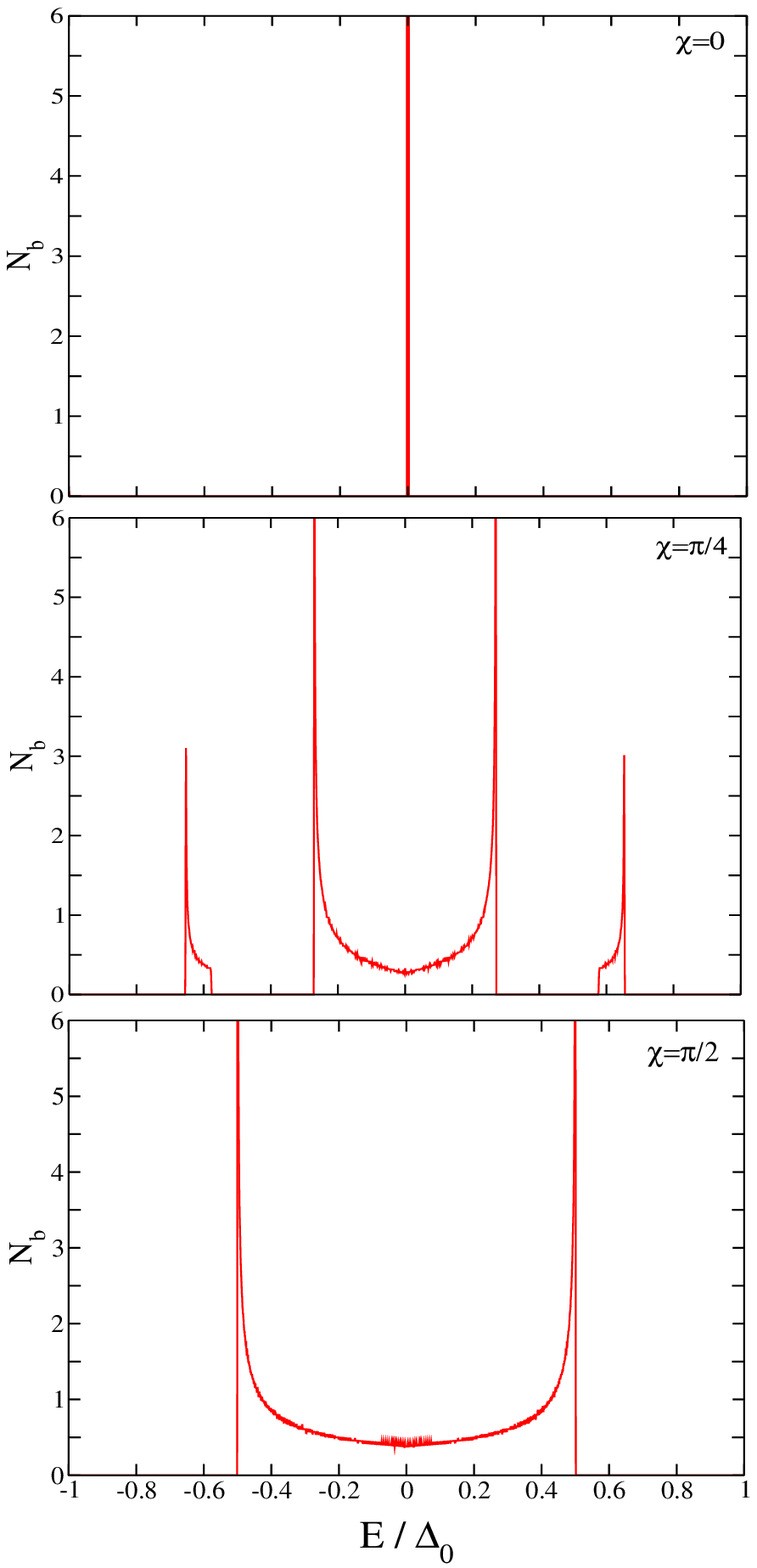}
\caption{(Color online) The energy dependence of the ABS contribution to the DOS (in arbitrary units) for domains with $\bm\eta\propto(1,0)$ and $(0,1)$, for different values of $\chi$. For $\chi=0$ the DOS has a delta-function singularity at $E=0$.}
\label{fig:DOS1}
\end{figure}

\subsection{$\bm{\eta}\propto(1,0)$ or $(0,1)$}

The order parameters in this nonchiral $p$-wave state have the form
\begin{equation}
\label{eta-nonchiral-1}
  \left.\begin{array}{ll}
        \bm{\eta}=\Delta_0(1,0),& x<0,\\ 
        \bm{\eta}=\Delta_0(0,1)e^{i\chi},& x>0,
        \end{array}\right.
\end{equation}
therefore, $\Delta_{-}(\theta)= \Delta_0 \cos \theta$, $\Delta_{+}(\theta)=\Delta_0 e^{i \chi} \sin\theta$, and
$$
  \beta=\frac{\cos\chi-\tan\theta}{\sin\chi}
$$
We calculate numerically the ABS spectra using Eq. (\ref{E_bound}) for different representative values of $\chi$ and present the results in 
Fig.~\ref{fig:E1}. Note that for some directions of semiclassical propagation the condition (\ref{cond}) is not satisfied and, therefore, there are no bound states. The case of $\chi=\pi/2$, when $\gamma_R=0$, is an exception: the ABS's
exist at all $\theta$. 

The ABS contribution to the DOS is plotted in Fig.~\ref{fig:DOS1}. For $\chi=0$, the DOS has a delta-function singularity at zero energy due to the flatness of the corresponding ABS dispersion. This can be understood as follows. 
At $\chi=0$, the gap functions on both sides of the DW are real, $\Delta_{-}(\theta)=\Delta_0\cos\theta$ and $\Delta_{+}(\theta)=\Delta_0\sin\theta$. If $\sin 2\theta<0$ then $\Delta_+$ and $\Delta_-$ have opposite signs.
In this case, the Andreev equations are equivalent to the Witten model of supersymmetric quantum mechanics and the ABS can be shown to have exactly zero energy, see Ref. \onlinecite{AGSY99} and the references therein and also Ref. \onlinecite{YH94}.
The condition $\mathrm{sgn}(\Delta_+\Delta_-)=-1$ is satisfied only at $\pi/2<\theta<\pi$ and at $3\pi/2<\theta<2\pi$, where the ABS energy is zero, as shown in the top panel of Fig. \ref{fig:E1}. 

Adding the phase $\chi$ makes the gap functions $\Delta_{\pm}$ intrinsically complex, in which case the analogy with the supersymmetric quantum mechanics breaks down and a simple physical explanation of the spectra
is no longer available. Our results show that as $\chi$ increases, the ABS bands acquire curvature and also the range of the ABS existence gets broader until, eventually, the internal gaps completely disappear for $\chi=\pi/2$, see the middle and bottom
panels of Fig. \ref{fig:E1}. At $0<\chi<\pi/2$, there are two sets of the van Hove singularities in the DOS, corresponding to the vanishing of the dispersion slope at the ABS band centers, 
and also two internal gaps, corresponding to the absence of the ABS at certain energies. Finally, at $\chi=\pi/2$, the four van Hove singularities merge into two.

\begin{figure}
\includegraphics[width=2.5in,height=4.5in, angle=0]{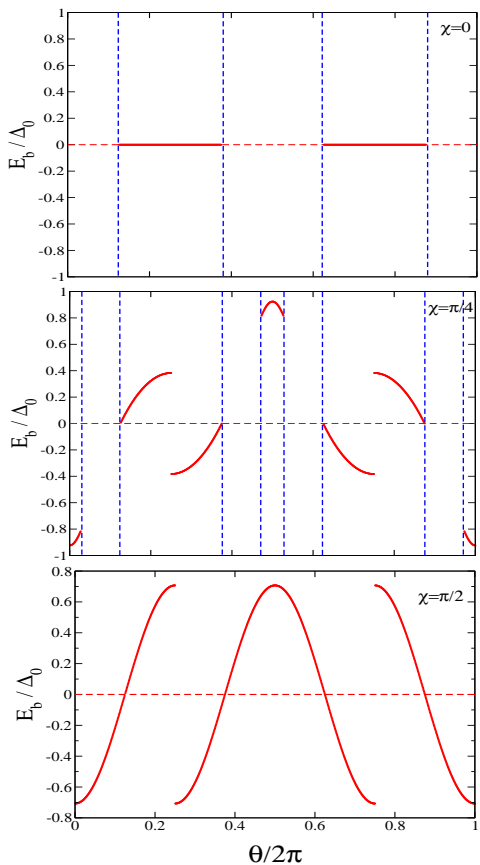}
\caption{(Color online) The variation of the ABS energy with the direction of quasiparticle propagation for domains with $\bm\eta\propto(1,\pm 1)$, for different values of $\chi$.
Regions bounded by blue vertical dotted lines and not containing any bound states correspond to the directions of propagation for which the condition (\ref{cond}) is not satisfied.}
\label{fig:E2}
\end{figure}
\vspace*{0.2in}
\begin{figure}
\includegraphics[width=2.5in,height=5.0in, angle=0]{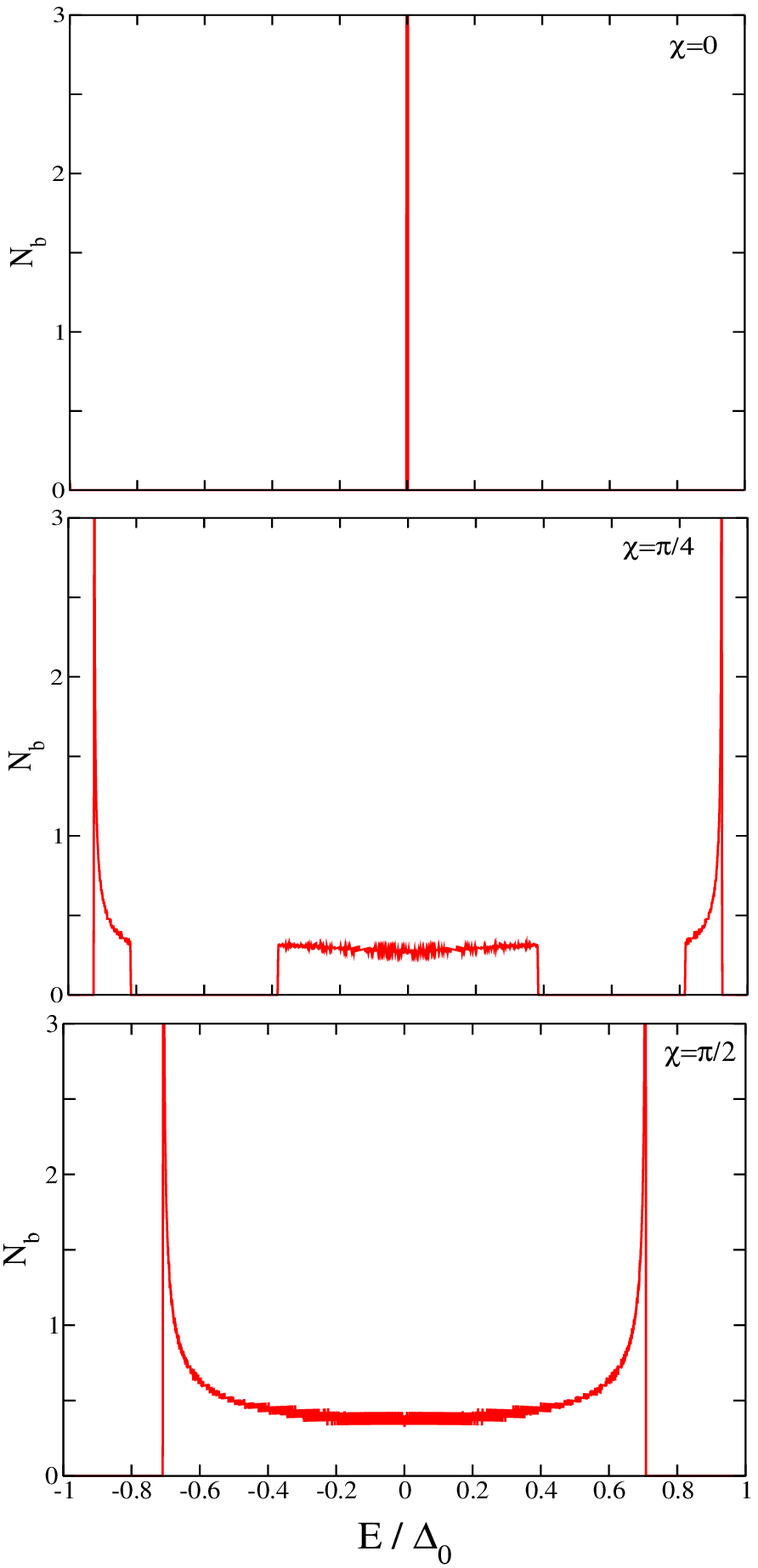}
\caption{(Color online) The energy dependence of the ABS contribution to the DOS (in arbitrary units) for domains with $\bm\eta\propto(1,\pm 1)$, for different values of $\chi$. For $\chi=0$ the DOS has a delta-function singularity at $E=0$.}
\label{fig:DOS2}
\end{figure}

\subsection{$\bm{\eta}\propto(1,\pm 1)$}

The order parameters in this nonchiral $p$-wave state have the form
\begin{equation}
\label{eta-nonchiral-2}
  \left.\begin{array}{ll}
        \bm{\eta}=\Delta_0(1,1),& x<0,\\ 
        \bm{\eta}=\Delta_0(1,-1)e^{i\chi},& x>0,
        \end{array}\right.
\end{equation}
therefore $\Delta_{-}(\theta)=\Delta_0 (\cos\theta + \sin\theta)$, $\Delta_{+}(\theta)= \Delta_0 e^{i \chi}(\cos \theta -\sin \theta) $, and 
$$
  \beta = \frac{\cos\chi-\tan(\pi/4-\theta)}{\sin\chi}.
$$
The ABS spectra and contribution to the DOS are calculated numerically for different representative values of $\chi$ and plotted in Figs.~\ref{fig:E2} and \ref{fig:DOS2}, respectively. 

Similarly to the previous case, the condition (\ref{cond}) 
is not satisfied for some values of $\theta$ (except at $\chi=\pi/2$), resulting in the spectra consisting of several disconnected parts and the gaps in the DOS. 
The van Hove singularities are again present, degenerating for $\chi=0$ into a delta-function singularity at zero energy. The latter's origin is explained by the fact that at $\chi=0$ the gap functions $\Delta_+$ and $\Delta_-$ are real 
and have opposite signs if $\cos 2\theta<0$, in which case the ABS has zero energy for $\pi/4\theta<3\pi/4$ and $5\pi/4\theta<7\pi/4$, as shown in the top panel of Fig. \ref{fig:E2}. At nonzero $\chi$, $\Delta_+$ and $\Delta_-$ become complex, which
results in the ABS bands acquiring dispersion and spreading to other angles of propagation until filling the entire angular range at $\chi=\pi/2$. 

\begin{figure}
\includegraphics[width=2.5in,height=4.5in, angle=0]{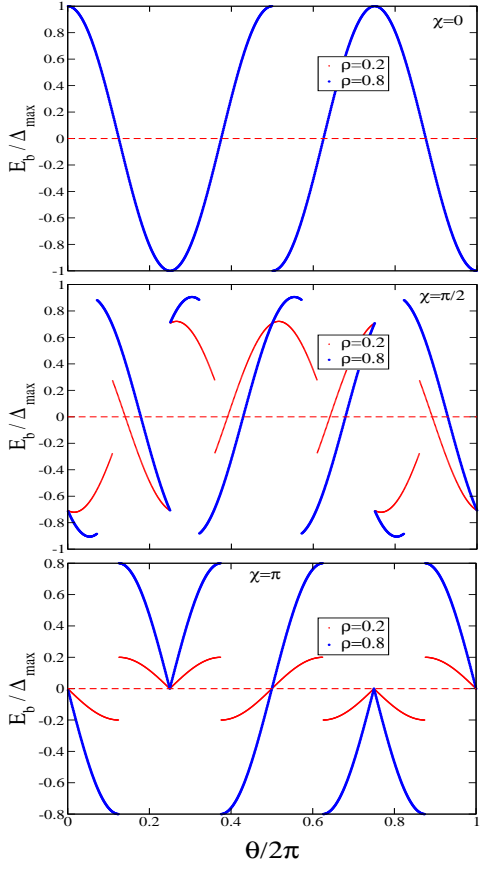}
\caption{(Color online) The variation of the ABS energy with the direction of quasiparticle propagation for the domains $d_{x^2-y^2}\pm id_{xy}$, for different values of $\chi$ and $\rho$. 
Thin red line is for $\rho=0.2$ and thick blue line is for $\rho=0.8$. For $\chi=0$ the curves superpose on each other.}
\label{fig:Ed2dxy}
\end{figure}
\vspace*{0.2in}
\begin{figure}
\includegraphics[width=2.5in,height=5.0in, angle=0]{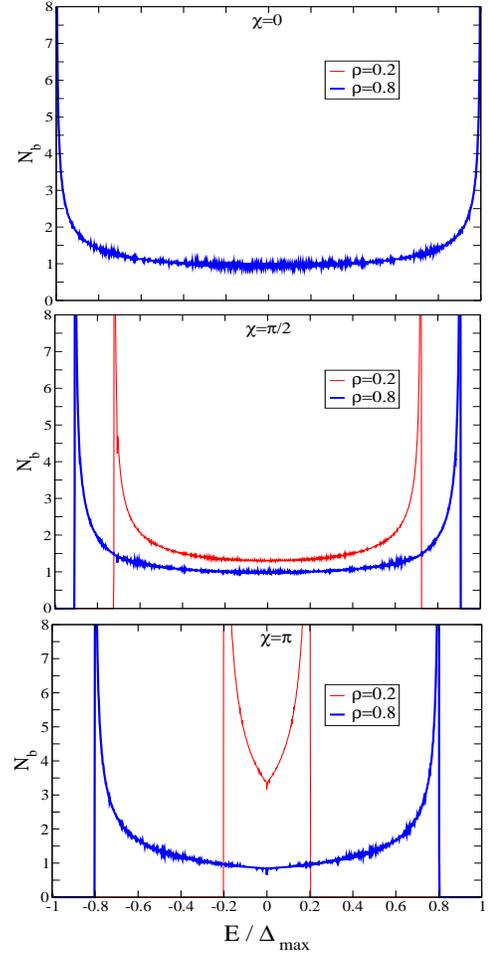}
\caption{(Color online) The energy dependence of the ABS contribution to the DOS (in arbitrary units) for the domains $d_{x^2-y^2}\pm id_{xy}$, for different values of $\chi$ and $\rho$. 
Thin red line is for $\rho=0.2$ and thick blue line is for $\rho=0.8$. For $\chi=0$ they superpose on each other.}
\label{fig:DOSd2dxy}
\end{figure}

\section{Mixed singlet states}
\label{sec:V}

In this section we discuss the ABS spectra for TRS-breaking mixtures of different singlet states. In all three cases, see Sec. \ref{sec:III}, the order parameters on both sides of a sharp DW can be written as
\be
\label{Delta-d-d}
  \left.\begin{array}{ll}
        (\psi_1,\psi_2)=(\Delta_1,i\Delta_2),& x<0, \\ 
        (\psi_1,\psi_2)=(\Delta_1,-i\Delta_2)e^{i\chi},& x>0,
        \end{array}\right.
\ee
where $\Delta_1,\Delta_2>0$ are constant amplitudes of the two singlet channels (without loss of generality, we assume that $\Delta_2\leq\Delta_1$) and $\chi$ is the common phase difference, see Appendix \ref{app: chi-singlet}.

\subsection{$d_{x^2-y^2} \pm id_{xy}$}

In this case, we obtain from Eqs. (\ref{Delta-pm-definition}), (\ref{Delta-singlet-gen}), and (\ref{varphi-dd}):
\begin{equation}
\left.\begin{array}{ll}
        \Delta_-(\theta) = \Delta_1\cos 2\theta + i \Delta_2 \sin 2\theta,\\
        \Delta_+(\theta) = \lp \Delta_1\cos 2\theta - i \Delta_2 \sin 2\theta\rp e^{i\chi}.
        \end{array}\right.
\end{equation}
We see that the gap magnitude is the same in both domains and given by 
$$
  |\Delta_+|=|\Delta_-|=\Delta_{bulk}(\theta)=\sqrt{\Delta_1^2\cos^22\theta+\Delta_2^2\sin^22\theta},
$$
which attains its maximum value, $\Delta_{max}=\Delta_1$, at $\theta=n\pi/2$, and its minimum value, $\Delta_{min}=\Delta_2$, at $\theta=(2n+1)\pi/4$, where $n=0,1,2,3$. Thus, the bulk gap has an anisotropic magnitude
without nodes, while the phases of the gap function are different in the two domains: $\Delta_-=\Delta_{bulk}e^{i\phi}$ and $\Delta_+=\Delta_{bulk}e^{i\chi}e^{-i\phi}$,
where 
$$
  \phi(\theta)=\arctan(\rho\tan 2\theta)
$$ 
and
\begin{equation}
\label{rho-def}
  \rho=\frac{\Delta_2}{\Delta_1}, \qquad 0\leq\rho\leq 1.
\end{equation}
From Eq. (\ref{beta-def}) we obtain:
\begin{equation}
\label{beta-dd}
  \beta=\tan\left(\phi-\frac{\chi}{2}\right),
\end{equation}
therefore, the expression (\ref{E_bound}) for the ABS spectrum takes the following form:
\bea
\label{Eb-dd}
  \frac{E_b(\theta)}{\Delta_{max}} = \sqrt{\rho^2 +(1-\rho^2)\cos^2 {2\theta}}\cos\left(\phi-\frac{\chi}{2}\right)\nonumber\\
  \times\,\mathrm{sgn}\left[\cos\theta\sin\left(\phi-\frac{\chi}{2}\right)\right],
\eea
which is shown in Fig. \ref{fig:Ed2dxy} for different representative values of $\rho$ and $\chi$. Although the ABS exist for all directions of semiclassical propagation, the spectra are not continuous, due to the presence of
the sign functions in Eq. (\ref{Eb-dd}). In some cases, relatively simple explicit expressions for the ABS energy can be obtained: for $\chi=0$, we have
$$
  \frac{E_b(\theta)}{\Delta_{max}} =\cos 2\theta\,\mathrm{sgn}(\sin\theta),
$$
which does not depend on $\rho$, and for $\chi=\pi$,
$$
  \frac{E_b(\theta)}{\Delta_{max}} =-\rho\sin 2\theta\,\mathrm{sgn}(\cos\theta\cos 2\theta).
$$
The DOS, shown in Fig. \ref{fig:DOSd2dxy}, has the van Hove singularities, corresponding to the vanishing of the dispersion slope. For $\chi=0$, the singularities occur at $|E|=1$, while for $\chi=\pi$ -- at $|E|=\rho$.

\begin{figure}
\includegraphics[width=2.5in,height=4.5in, angle=0]{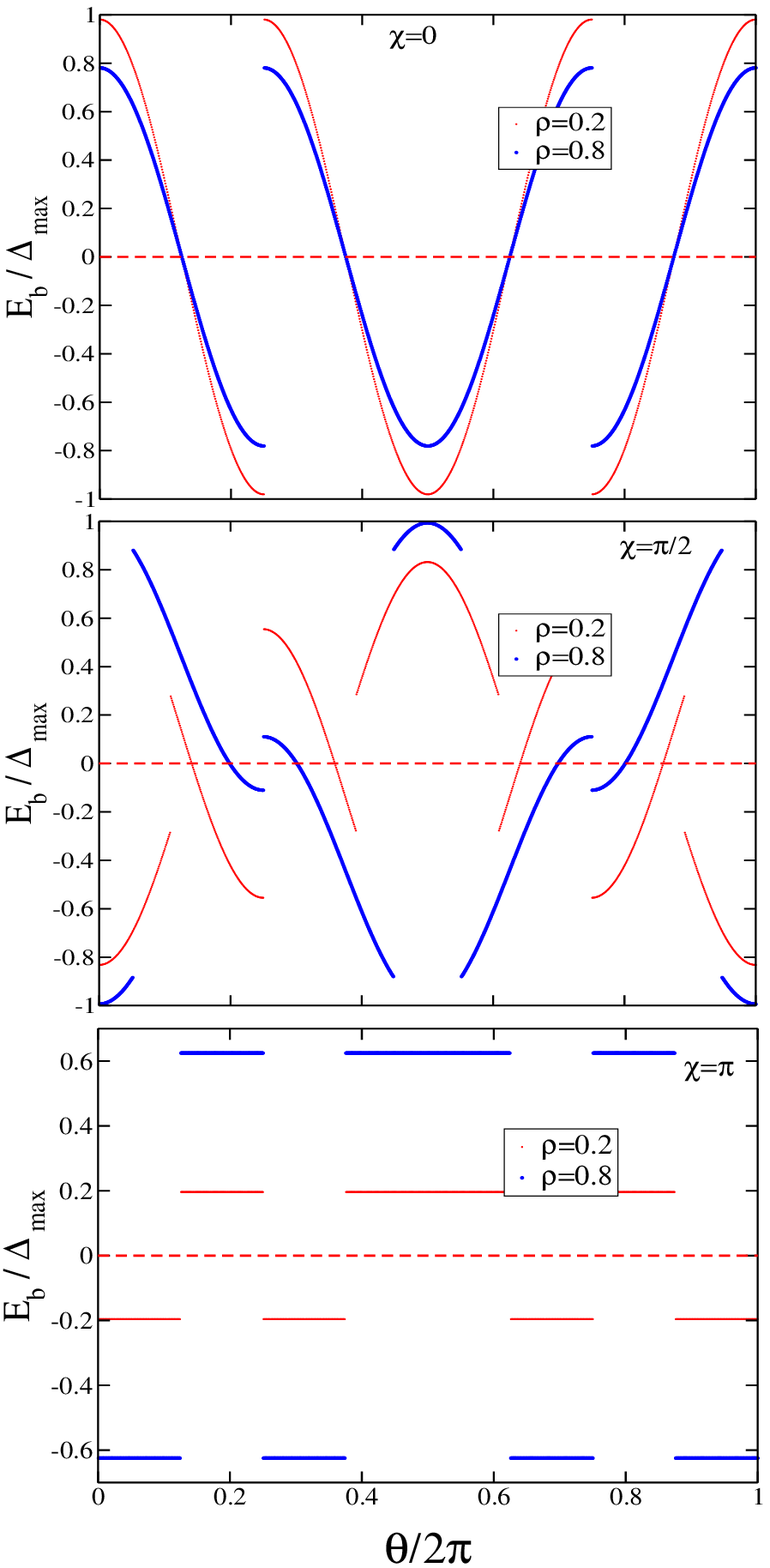}
\caption{(Color online) The variation of the ABS energy with the direction of quasiparticle propagation for the domains $d_{x^2-y^2}\pm is$, for different values of $\chi$ and $\rho$. 
Thin red line is for $\rho=0.2$ and thick blue line is for $\rho=0.8$.}
\label{fig:Ed2s}
\end{figure}
\vspace*{0.2in}
\begin{figure}
\includegraphics[width=2.5in,height=5.0in, angle=0]{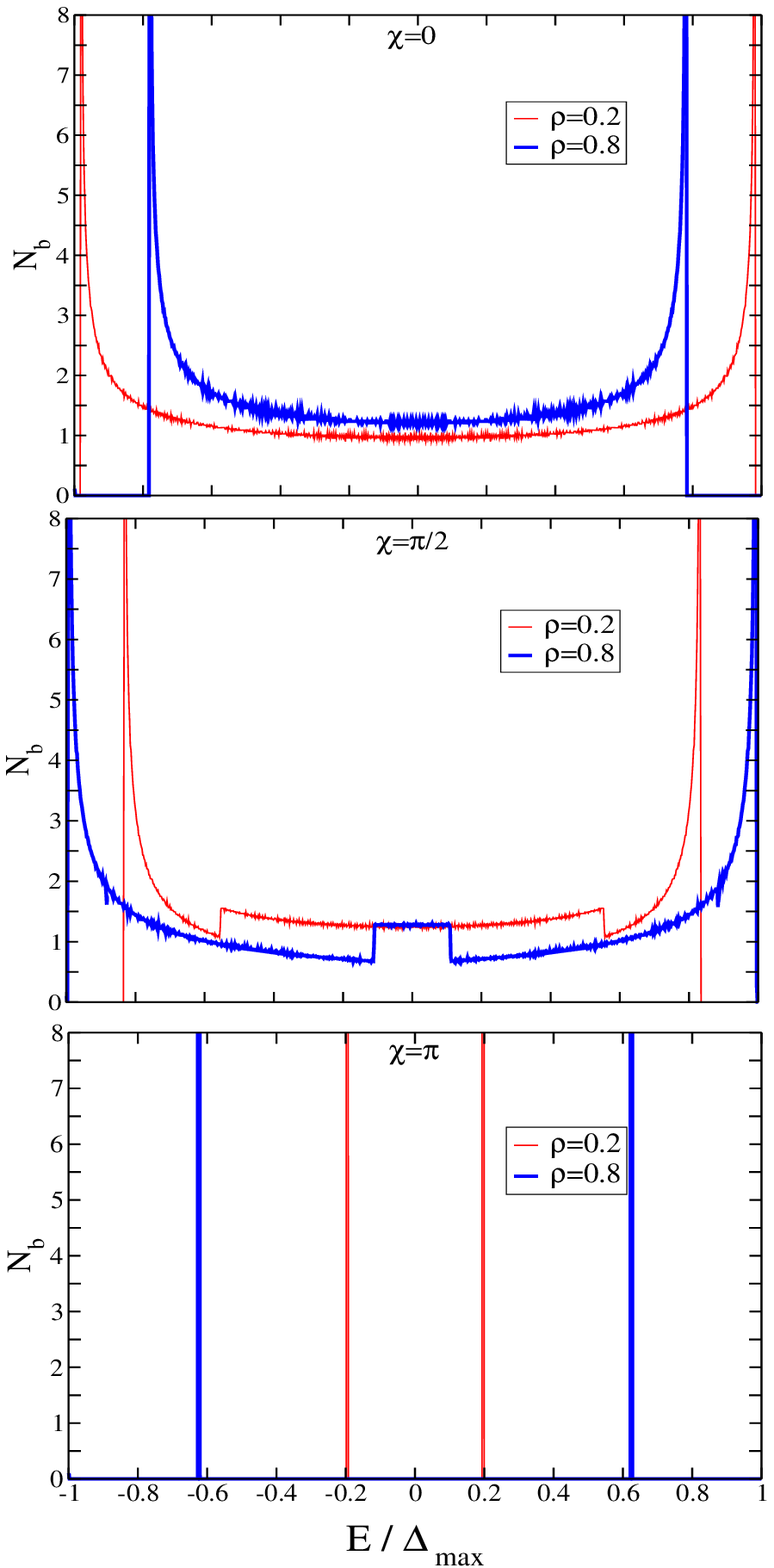}
\caption{(Color online) The energy dependence of the ABS contribution to the DOS (in arbitrary units) for the domains $d_{x^2-y^2}\pm is$, for different values of $\chi$ and $\rho$. 
Thin red line is for $\rho=0.2$ and thick blue line is for $\rho=0.8$. For $\chi=\pi$ the DOS has delta-function singularities.}
\label{fig:DOSd2s}
\end{figure}

\subsection{$d_{x^2-y^2}\pm is$}

According to Eqs. (\ref{Delta-pm-definition}), (\ref{Delta-singlet-gen}), and (\ref{varphi-d2s}), the superconducting domains are described by the following gap functions:
\begin{equation}
\left.\begin{array}{ll}
        \Delta_-(\theta) = \Delta_1\cos 2\theta + i \Delta_2,\\
        \Delta_+(\theta) = \lp \Delta_1\cos 2\theta - i \Delta_2\rp e^{i\chi}.
        \end{array}\right.
\end{equation}
The gap magnitudes are the same in both domains: 
$$
  |\Delta_+|=|\Delta_-|=\Delta_{bulk}(\theta)=\sqrt{\Delta_1^2\cos^22\theta+\Delta_2^2}.
$$
The bulk gap magnitude is anisotropic, without nodes.
It attains its maximum value, $\Delta_{max}=\sqrt{\Delta_1^2+\Delta_2^2}$, at $\theta=n\pi/2$, and the minimum value, $\Delta_{min}=\Delta_2$, at $\theta=(2n+1)\pi/4$, where $n=0,1,2,3$.

The parameter $\beta$ is given by Eq. (\ref{beta-dd}), where 
$$
  \phi=\arctan\left(\frac{\rho}{\cos 2\theta}\right)
$$
and $\rho$ is given by Eq. (\ref{rho-def}). Therefore, we obtain for the ABS spectrum:
\bea
\label{Eb-d2-s}
 \frac{E_b(\theta)}{\Delta_{max}}=\sqrt{\frac{\rho^2 + \cos^2 {2\theta}}{\rho^2 +1}}\cos\left(\phi-\frac{\chi}{2}\right)\nonumber\\
  \times\,\mathrm{sgn}\left[\cos\theta\sin\left(\phi-\frac{\chi}{2}\right)\right],
\eea
which is plotted in Fig.~\ref{fig:Ed2s}. One can see that the ABS exist for all $\theta$, but their spectra are not continuous. The DOS is shown in Fig. \ref{fig:DOSd2s}. 
In particular, in the case of $\chi=\pi$, we find from Eq. (\ref{Eb-d2-s}) that $|E_b|/\Delta_{max}=\rho/\sqrt{\rho^2+1}$, i.e. does not depend on $\theta$, which results in the delta-function
singularities in the DOS.

\begin{figure}
\includegraphics[width=2.5in,height=4.5in, angle=0]{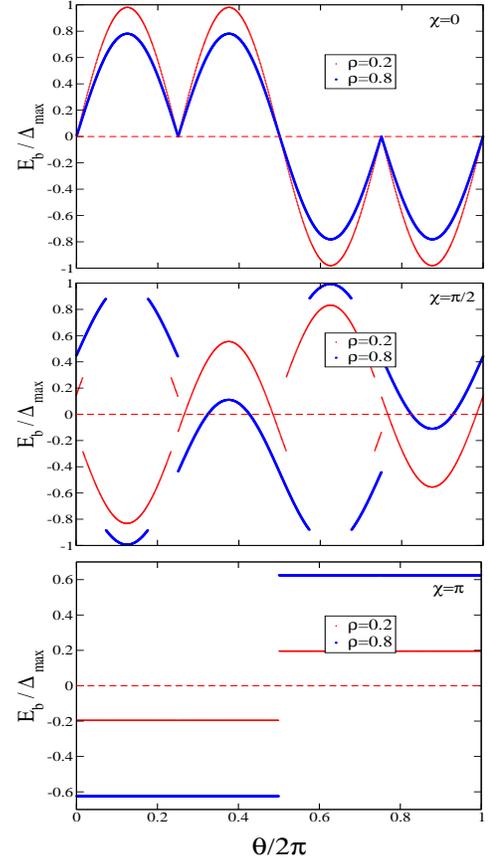}
\caption{(Color online) The variation of the ABS energy with the direction of quasiparticle propagation for the domains $d_{xy} \pm is$, for different values of $\chi$ and $\rho$. 
Thin red line is for $\rho=0.2$ and thick blue line is for $\rho=0.8$.}
\label{fig:Edxys}
\end{figure}
\vspace*{0.2in}
\begin{figure}
\includegraphics[width=2.5in,height=5.0in, angle=0]{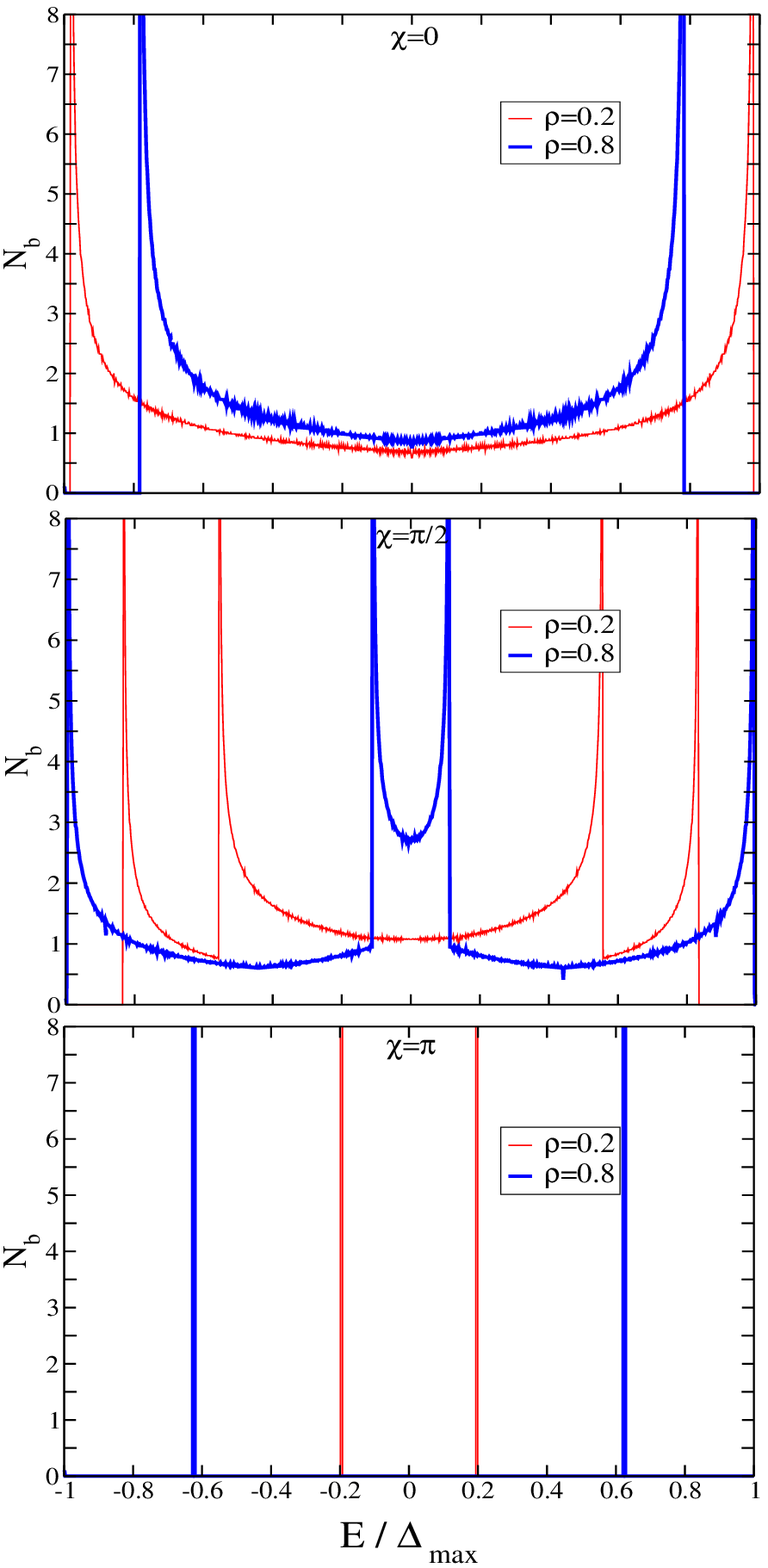}
\caption{(Color online) The energy dependence of the ABS contribution to the DOS (in arbitrary units) for the domains $d_{xy}\pm is$, for different values of $\chi$ and $\rho$. 
Thin red line is for $\rho=0.2$ and thick blue line is for $\rho=0.8$. For $\chi=\pi$ the DOS has delta-function singularities.}
\label{fig:DOSdxys}
\end{figure}

\subsection{$d_{xy}\pm is$ }

According to Eqs. (\ref{Delta-pm-definition}), (\ref{Delta-singlet-gen}), and (\ref{varphi-dxys}), the superconducting domains are described by the following gap functions:
\begin{equation}
\left.\begin{array}{ll}
        \Delta_-(\theta) = \Delta_1\sin 2\theta + i \Delta_2,\\
        \Delta_+(\theta) = \lp \Delta_1\sin 2\theta - i \Delta_2\rp e^{i\chi}.
        \end{array}\right.
\end{equation}
The gap magnitudes are the same in both domains: 
$$
  |\Delta_+|=|\Delta_-|=\Delta_{bulk}(\theta)=\sqrt{\Delta_1^2\sin^22\theta+\Delta_2^2}.
$$
The bulk gap magnitude is anisotropic, without nodes.
It attains its maximum value, $\Delta_{max}=\sqrt{\Delta_1^2+\Delta_2^2}$, at $\theta=(2n+1)\pi/4$, where $n=0,1,2,3$, and the minimum value, $\Delta_{min}=\Delta_2$, at $\theta=n\pi/2$, where $n=0,1,2,3$. 

The parameter $\beta$ is given by Eq. (\ref{beta-dd}), where 
$$
  \phi=\arctan\left(\frac{\rho}{\sin 2\theta}\right)
$$  
and $\rho$ is given by Eq. (\ref{rho-def}). The ABS spectrum has the following form:
\bea
\label{Eb-dxy-s}
  \frac{E_b(\theta)}{\Delta_{max}}=\sqrt{\frac{\rho^2 + \sin^2 {2\theta}}{\rho^2 +1}} \cos\left(\phi-\frac{\chi}{2}\right)\nonumber\\
  \times\,\mathrm{sgn}\left[\cos\theta\sin\left(\phi-\frac{\chi}{2}\right)\right],
\eea
which is shown in Fig.~\ref{fig:Edxys}. Again, the ABS exist for all $\theta$ and their spectra are not continuous. The DOS is shown in Fig. \ref{fig:DOSdxys}. 
In particular, in the case of $\chi=\pi$, we find from Eq. (\ref{Eb-dxy-s}) that $|E_b|/\Delta_{max}=\rho/\sqrt{\rho^2+1}$, i.e. does not depend on $\theta$, which results in the delta-function
singularities in the DOS.

\section{Conclusion}
\label{sec:VI}

We have shown that the DWs separating degenerate ground states in unconventional superconductors can trap fermionic quasiparticles and create Andreev bound states with energies inside the bulk gap. Using 
as an example a quasi-2D tetragonal superconductor, we have studied both chiral and nonchiral $p$-wave states, as well as the mixed singlet states of $d+id$ and $d+is$ symmetries. Time reversal symmetry is broken 
in each case, except the nonchiral $p$-wave states.

We derived a general expression for the ABS energy in the semiclassical approximation, for a sharp DW model. The bound states exist only for certain directions of the quasiparticle propagation, 
and there is only one bound state for each ``allowed'' direction. The ABS spectrum is highly anisotropic, forming the ABS ``bands'' as the direction of the Fermi wavevector varies, which results 
in some prominent features in the quasiparticle DOS, such as internal gaps, abrupt steps, and the van Hove singularities. 

Some of the DOS features are specific to certain pairing symmetries and can help to distinguish them. 
For instance, the ABS spectrum in the chiral $p$-wave state has neither internal gaps nor delta-function peaks, in contrast to the nonchiral ones. 
In the latter cases, the origin of the flat zero-energy ABS bands and the associated delta-function peaks in the DOS is explained by the sign change of the real gap function along the 
semiclassical trajectory. In the mixed singlet cases, the spectra of $d+is$ states can have four delta-function peaks at nonzero energies due to the Andreev bands with flat dispersion, which never happens in the $d+id$ state. 
We hope that the abovementioned features can be detected in tunneling experiments, thus providing a direct evidence of the DW presence and shedding light on the pairing symmetry of unconventional superconductors.

\appendix

\section{Calculation of $\chi$ for the triplet states}
\label{app: chi-triplet}

Superconducting current can be obtained from the gradient terms in the Ginzburg-Landau free energy density. For a $p$-wave triplet order parameter in a tetragonal crystal the latter have the following form:\cite{Mineev}
\bea
\nonumber
  F_{grad}= K_{1}(\nabla_i\eta_{j})^* (\nabla_i\eta_{j}) + K_{2}(\nabla_i\eta_{i})^* (\nabla_{j}\eta_{j})\\
    +K_{3}(\nabla_i\eta_{j})^* (\nabla_{j}\eta_{i}) + K_{4}(\nabla_i\eta_{i})^* (\nabla_i\eta_{i}),
\eea
where the summation over repeated indices, $i,j=x,y$, is implied. Replacing the gradients by the covariant derivatives, $\bm{\nabla}\to\bm{\nabla}+2i\bm{A}$, and varying with respect to the vector potential $\bm{A}$, we obtain:
\bea
\nonumber
  j_{i}=2\,\mathrm{Im}\,( K_{1} \eta_{j}^* \nabla_{i} \eta_{j} +  K_{2} \eta_{i}^* \nabla_{j} \eta_{j}\\ \nonumber 
  +K_{3} \eta_{j}^* \nabla_{j} \eta_{i} + K_{4} \eta_{i}^* \nabla_{i} \eta_{i}).
\eea

The order parameter components can be written in the following general form:
\be
\eta_1 = \Delta_0 f_1(x)e^{i\phi(x)},\quad\eta_2 = \Delta_0 f_2(x)e^{i\phi(x)-i\gamma(x)},
\label{eta12}
\ee
where $f_{1,2}$ are dimensionless amplitudes, $\phi$ is the common (or Josephson) phase, and $\gamma$ is the relative phase. From these expressions we 
calculate the components of the supercurrent as follows:
\be
  j_x = 2 \Delta_0^2 \lp K_{1234} f_{1}^2 + K_{1} f_{2}^2 \rp \nabla_{x} \phi - 2 \Delta_0^2 K_{1} f_{2}^2\nabla_{x} \gamma,
\label{Jx}
\ee
where $K_{1234} = K_1 + K_2 + K_3 + K_4$, and
\bea
\nonumber
  j_y =2 \Delta_0^2 [ \sin \gamma \lp K_2 f_2 \nabla_x f_1 - K_3 f_1 \nabla_x f_2\rp\\
  +f_1 f_2 \cos \gamma [(K_2 + K_3) \nabla_x \phi - K_3 \nabla_x\gamma].
\label{jy-triplet}
\eea
From the current conservation we get $j_x=\mathrm{const}$, where the constant is determined by the boundary conditions and can be set to zero. This gives us a relation between the phase gradients:
$$
\nabla_x \phi= \frac{K_1 f_2^2}{K_{1234} f_1^2 + K_1 f_2^2} \nabla_x \gamma.
$$
Therefore,
\bea
\nonumber
  \chi &\equiv& \phi(+ \infty)-\phi(-\infty)\\
    &=& \int_{-\infty}^{\infty}dx\, \frac{K_1 f_2^2}{K_{1234}f_1^2+K_1 f_2^2} \nabla_x\gamma.
    \label{chi-general}
\eea
We see that the parameter $\chi$ depends on the microscopic characteristics of the system, through the coefficients $K_i$, as well as on the DW structure, described by $f_{1,2}(x)$ and $\gamma(x)$, which are model-specific. 
Below we calculate $\chi$ for different triplet states in a simple approximation in which the order amplitudes are constant, either globally, i.e. at all $x$, or separately at $x>0$ and $x<0$. 

In the chiral case with $\bm{\eta}\propto(1, \pm i)$, we assume that $f_1=f_2=1$ at all $x$, see Ref. \onlinecite{Volovik}. Since $\gamma(\pm \infty) = \pm\pi/2$, we can calculate the integral on
the right-hand side of Eq. (\ref{chi-general}), with the following result:
\begin{equation}
\label{chi-final}
  \chi = \frac{K_1}{K_{1234}+K_1} \pi.
\end{equation}
For the nonchiral state with $\bm{\eta}\propto (1, \pm 1)$ we have $f_1=f_2=1$, but $\gamma(-\infty)=0$ and $\gamma(+\infty)=\pi$. From Eq. (\ref{chi-general}), we obtain the same expression (\ref{chi-final}) for $\chi$.  
For the nonchiral state with $\bm{\eta}\propto (1,0)$ or $(0,1)$, we have $f_{1}=\Theta(-x)$ and $f_{2}=\Theta(x)$, where $\Theta(x)$ is the Heaviside step function. In this case, the current conservation is satisfied if
both $\phi$ and $\gamma$ take arbitrary constant values throughout the system, which means that the order parameter can be written as
$\eta_1=\Delta_0\Theta(-x),\quad \eta_2=\Delta_0\Theta(x)e^{i\chi}$, where the value of $\chi$ is arbitrary in our approximation. Using Eq. (\ref{jy-triplet}) one can show that the net current along the DW is 
nonzero in the chiral state, but vanishes in both nonchiral states.

\section{Calculation of $\chi$ for the singlet states}
\label{app: chi-singlet}

As an example, we consider only the $d_{x^2-y^2}\pm is$ state, the calculation for $d_{xy}\pm is$ and $d_{x^2-y^2}\pm id_{xy}$ states being similar.
The gradient part of the Ginzburg-Landau free energy density has the form\cite{Ren-WWL}
\bea
\nonumber
  && F_{grad}=K_{1}|\bm{\nabla}\psi_1|^2+K_{2}|\bm{\nabla}\psi_2|^2\\
  &&\quad +K_{3}[(\nabla_{y}\psi_1)^* (\nabla_{y}\psi_2)-(\nabla_{x}\psi_1)^*(\nabla_{x}\psi_2) + c.c.]\qquad
\eea
(for the $d+id$ state the $K_3$ term is absent). We assume that the order parameters $\psi_{1,2}$ depend only on $x$. Following the same procedure as in Appendix \ref{app: chi-triplet}, we calculate the current across the DW and along the DW,
with the following results: 
\bea
\nonumber
  j_x= 2\, \mathrm{Im}\, [ K_{1} (\nabla_{x}\psi_{1}^*)\psi_{1} + K_{2} (\nabla_{x}\psi_{2}^*)\psi_{2} \\
    +K_{3}\lp\psi_{1}^*\nabla_{x}\psi_{2} + \psi_{2}^*\nabla_{x}\psi_{1}\rp]
\eea
and $j_y = 0$. 

Representing the order parameters in the form
\be
  \psi_1=\Delta_1(x)e^{i \phi(x)},\ \psi_2=\Delta_2(x)e^{i\phi(x)-i\gamma(x)},
\ee
where $\Delta_{1,2}$ are the amplitudes, $\phi$ is the common phase, and $\gamma$ is the relative phase, we find from the current conservation that the phase gradients are related:
\be
  \nabla_{x} \phi = \frac{K_{2}\Delta_2^2 + K_{3} \Delta_1\Delta_2\cos \gamma}{K_{1} \Delta_1^2 + K_{2} \Delta_2^2 + 2 K_{3} \Delta_1\Delta_2\cos \gamma} \nabla_{x} \gamma.
\ee
Therefore,
\bea
\nonumber
 \chi & \equiv & \phi(+ \infty)-\phi(-\infty)\\ \nonumber
 &=& \int_{-\infty}^{\infty} dx\,\frac{K_{2}\Delta_2^2 + K_{3} \Delta_1\Delta_2\cos \gamma}{K_{1} \Delta_1^2 + K_{2} \Delta_2^2 + 2 K_{3}\Delta_1 \Delta_2\cos \gamma}\nabla_x\gamma.\\
\eea
The relative phase satisfies the following boundary conditions: $\gamma(\pm \infty) = \pm\pi/2$,

Assuming constant amplitudes $\Delta_1$ and $\Delta_2$, we have
\be
\label{chi-mixed-exact}
  \chi =\frac{1}{2} \int_{-\pi/2}^{\pi/2} \frac{c_1 + \cos \gamma}{c_2 + \cos \gamma} d\gamma,
\ee
where 
$$
  c_1= \frac{K_{2}\Delta_2}{K_{3}\Delta_1},\qquad c_2=\frac{K_{1}\Delta_1^2+K_{2}\Delta_2^2}{2K_{3}\Delta_1\Delta_2}.
$$ 
The integral in Eq. (\ref{chi-mixed-exact}) can be solved analytically and we finally obtain:
$$
  \chi = \frac{\pi}{2} + \frac{2(c_1-c_2)}{\sqrt{c_2^2 - 1}} \arctan\sqrt{\frac{c_2-1}{c_2+1}},
$$
for $c_2>1$, and
$$
  \chi = \frac{\pi}{2} + \frac{(c_1-c_2)}{\sqrt{c_2^2 - 1}} \ln \lp \frac{c_2-1-\sqrt{c_2^2 - 1}}{c_2-1+\sqrt{c_2^2 - 1}}\rp,
$$
for $c_2<1$.

\end{document}